\documentclass[twocolumn,times]{aastex61}

\shorttitle{Observations of Current Sheet Formation}
\shortauthors{Warren et al.}

\journalinfo{To be submitted to ApJ; This version: \today}

\begin{document}

%% ------------------------------------------------------------------------------------------
%% --- TITLE PAGE ---------------------------------------------------------------------------
%% ------------------------------------------------------------------------------------------

\title{Spectroscopic Observations of Current Sheet Formation and Evolution}

\correspondingauthor{Harry P. Warren}
%\email{harry.warren@nrl.navy.mil}

\author{Harry P. Warren}
\affil{Space Science Division, Naval Research Laboratory, Washington, DC 20375}

\author{David H. Brooks}
\affil{College of Science, George Mason University, 4400 University Drive, Fairfax, VA 22030 USA}
\affil{Hinode Team, ISAS/JAXA, 3-1-1 Yoshinodai, Chuo-ku, Sagamihara, Kanagawa 252-5210, Japan}

\author{Ignacio Ugarte-Urra}
\affil{Space Science Division, Naval Research Laboratory, Washington, DC 20375}

\author[0000-0003-4739-1152]{Jeffrey W. Reep}
\affil{National Research Council Postdoctoral Fellow, Space Science Division, Naval Research
  Laboratory, Washington, DC 20375, USA}

\author{Nicholas A. Crump}
\affil{Space Science Division, Naval Research Laboratory, Washington, DC 20375}

\author{George A. Doschek}
\affil{Space Science Division, Naval Research Laboratory, Washington, DC 20375}

%% ------------------------------------------------------------------------------------------
%% --- ABSTRACT -----------------------------------------------------------------------------
%% ------------------------------------------------------------------------------------------

\begin{abstract}
We report on the structure and evolution of a current sheet that formed in the wake of an eruptive
X8.3 flare observed at the west limb of the Sun on September 10, 2017. Using observations from the
EUV Imaging Spectrometer (EIS) on \textit{Hinode} and the Atmospheric Imaging Assembly (AIA) on the
\textit{Solar Dynamics Observatory} (\textit{SDO}), we find that plasma in the current sheet
reaches temperatures of about 20\,MK and that the range of temperatures is relatively narrow. The
highest temperatures occur at the base of the current sheet, in the region near the top of the
post-flare loop arcade. The broadest high temperature line profiles, in contrast, occur at the
largest observed heights. Further, line broadening is strong very early in the flare and diminishes
over time. The current sheet can be observed in the AIA 211 and 171 channels, which have a
considerable contribution from thermal bremsstrahlung at flare temperatures. Comparisons of the
emission measure in these channels with other EIS wavelengths and AIA channels dominated by Fe line
emission indicate a coronal composition and suggest that the current sheet is formed by the heating
of plasma already in the corona. Taken together, these observations suggest that some flare heating
occurs in the current sheet while additional energy is released as newly reconnected field lines
relax and become more dipolar.
\end{abstract}

\keywords{Sun: corona, Sun: flare}

%% ------------------------------------------------------------------------------------------
%% --- BODY ---------------------------------------------------------------------------------
%% ------------------------------------------------------------------------------------------

\section{Introduction}

It is widely thought that eruptive solar flares are powered by the energy released during magnetic
reconnection \citep[e.g.,][]{shibata1995}. The standard thinking is that turbulent photospheric
motions lead to the twisting and braiding of magnetic fields in the corona and that this
topological complexity is ultimately dissipated through magnetic reconnection. The process of how
energy stored in the magnetic field is transferred to the plasma, however, is poorly
understood. For example, it is not clear if energy transport occurs through the formation of shocks
from field lines retracting from the reconnection region, the acceleration of particles in the
current sheet, the dissipation of waves, or through some combination of these, and perhaps other
mechanisms. The lack of progress on these issues stems from the difficulty of observing magnetic
reconnection directly. Theory suggests that it occurs on very small spatial scales
\citep[e.g.][]{zweibel2009}. Furthermore, the plasma processed through the current sheet is from
the relatively tenuous corona. This combination of small volumes and low densities should lead to
low emission measure and make it challenging to observe.

In this paper we report on the structure and evolution of a current-sheet-like feature that formed
in the wake of a solar eruption and flare that occurred on the west limb of the Sun on September
10, 2017. There is no commonly accepted observational signature of a current sheet, but, as we will
see, this event has all of the expected features. The eruption begins with the heating of a
filament low in the corona. As the filament erupts it forms a cavity that expands as it moves
outward. Bright, high-temperature emission develops at the base of the eruption and eventually
evolves into a classic post-flare loop arcade with a cusp-like shape at its top. Finally, a long,
narrow, linear structure forms from the top of the cusp and follows the path of the erupting
filament. It is this feature that we assume is the current sheet where magnetic reconnection is
occurring.

Here we discuss observations of the event from the EUV Imaging Spectrometer (EIS;
\citealt{culhane2007}) on \textit{Hinode} \citep{kosugi2007} and the Atmospheric Imaging Assembly
(AIA; \citealt{lemen2012}) on the \textit{Solar Dynamics Observatory} (\textit{SDO}). At the time
of the eruption EIS was executing a flare watch study consisting of short exposures over a
relatively wide field of view that extended up to about $1.15R_{\odot}$. We use the ratio of the
EIS \ion{Fe}{24} 255.10\,\AA\ to \ion{Fe}{23} 263.76\,\AA\ lines to infer temperatures of
15--20\,MK in the current sheet. The highest temperatures occur near the base of the current sheet
in the cusp-shaped feature at the top of the post-flare loop arcade and decline with height. Many
previous studies had suggested elevated temperatures in this region relative to the brightest
emission in the flare arcade \citep[e.g.,][]{tsuneta1997,warren1999}. Here we report the first
temperature measurements that have extended into the current sheet just above the
arcade. Comparisons with the AIA observations in this region indicate that the distribution of
temperatures is relatively narrow, as it appears bright in the 193 and 131 channels that are
dominated by high temperature emission lines and dim in the channels dominated by lower temperature
lines. This is confirmed by a formal inversion of the observed intensities to compute the
differential emission measure distribution.

Many previous studies of non-thermal line broadening in high temperature emission lines have
utilized spatially unresolved observations
\citep[e.g.,][]{doschek1980,antonucci1984,mariska1993}. These studies have found that the largest
non-thermal velocities generally occur early in a flare, sometime even before any hard X-ray
emission is detected \citep[e.g.,][]{alexander1998,harra2001}. We use spatially resolved
observations of the EIS \ion{Fe}{24} 192.04\,\AA\ line to measure high temperature line
broadening. We find that the strongest non-thermal velocities occur early in the flare, consistent
with earlier results. Further, we find that non-thermal broadening increases with height in the
current sheet.

The current sheet can be observed in the AIA 211 and 171 channels, which do not have strong
contributions from emission lines at flare temperatures \citep{odwyer2010}. Thus the emission from
the current sheet in these channels is likely to be from thermal bremsstrahlung. Comparisons with
the emission from other EIS wavelengths and AIA channels dominated by Fe line emission can be used
to infer elemental abundances. We find that the observations are consistent with a coronal
composition in the current sheet.

Several previous eruptive flares have shown evidence for a current sheet, but such observations are
relatively rare. Almost all previous spectroscopic observations appear to have been taken with the
Ultraviolet Coronagraph Spectrometer (UVCS; \citealt{kohl1995}) on the \textit{Solar and
  Heliospheric Observatory} (\textit{SoHO}). \citet{ciaravella2002}, \citet{ko2003},
\citet{ciaravella2008}, and \citet{schettino2010} report on UVCS observations of four different
events, all observed at heights above about $1.5R_{\odot}$. Some spectroscopic observations from
EIS near the base of the current sheet have been reported by \cite{hara2008}. The properties of the
current sheet have also been studied using imaging observations
\citep[e.g.,][]{savage2010,seaton2017,zhu2016}. Our results are generally consistent with these
previous studies and we will discuss this in detail in the final section of the paper.
\section{Observations}

\begin{figure}[t!]
  \centerline{\includegraphics[width=0.5\textwidth]{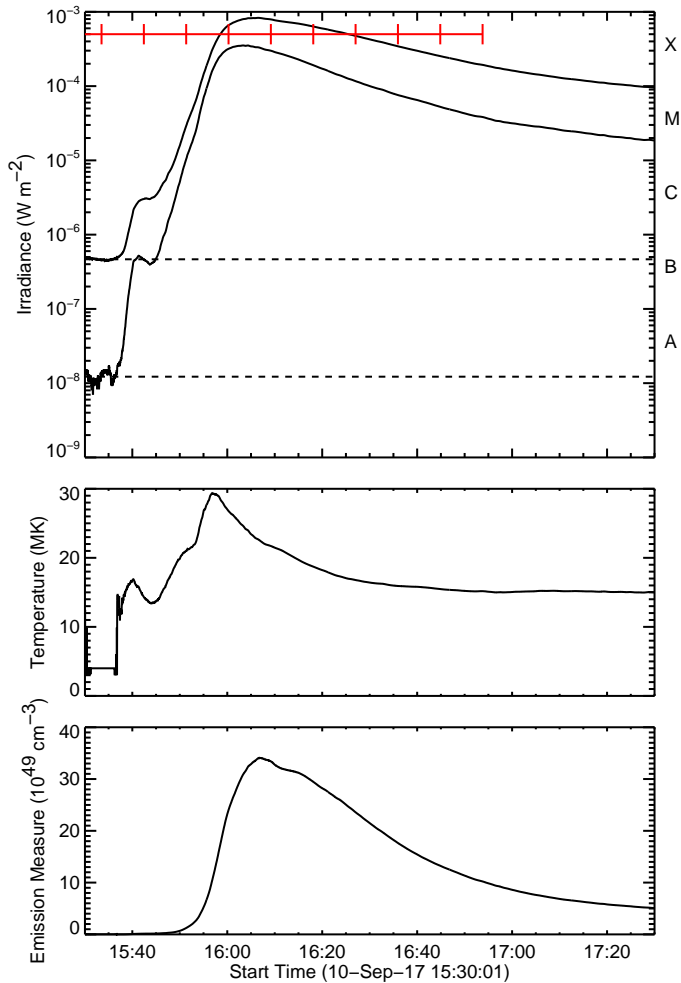}}
  \caption{\textit{GOES} soft X-ray observations for the X8.3 flare that occurred on 10 September
    2017 near 16\,UT. \textit{GOES} indicates the presence of relatively high temperature emission
    in the flare, but the spatial distribution of this hot plasma cannot be determined from these
    integrated observations. Top: \textit{GOES} light curves for the 1--8 and
    0.5--4\,\AA\ channels. Dashed lines indicate the assumed background levels. The red ticks at
    the top of the plot indicate the start and stop time for each of the EIS rasters taken during
    this period. Middle and bottom: The temperature and emission measure derived from the
    \textit{GOES} 0.5--4 to 1--8\,\AA\ ratio.}
  \label{fig:goes}
\end{figure}

In this paper we focus on the spatially resolved EIS and AIA observations of the September 10, 2017
X flare. We begin, however, with a discussion of some context soft X-ray irradiance data from the
X-Ray Sensor (XRS; see, for example, \citealt{garcia1994}) on the \textit{Geostationary Operational
  Environmental Satellite} (\textit{GOES}). Light curves from the 1--8 and 0.5--4\,\AA\ channels
are shown in Figure~\ref{fig:goes}. The peak in the 1--8 channel occurred at 16:06:28 at a level of
X8.3 ($8.3\times10^4$\,W~m$^{-2}$). The ratio of the two XRS channels can be used to infer a
response-weighted average temperature \cite[e.g.,][]{white2005}. This analysis is also shown in
Figure~\ref{fig:goes} and indicates a peak temperature of about 29\,MK at 15:56:49. The spatial
distribution of this hot plasma, however, cannot be determined from these integrated observations.

\begin{figure*}[t!]
  \centerline{\includegraphics[angle=90,width=\textwidth]{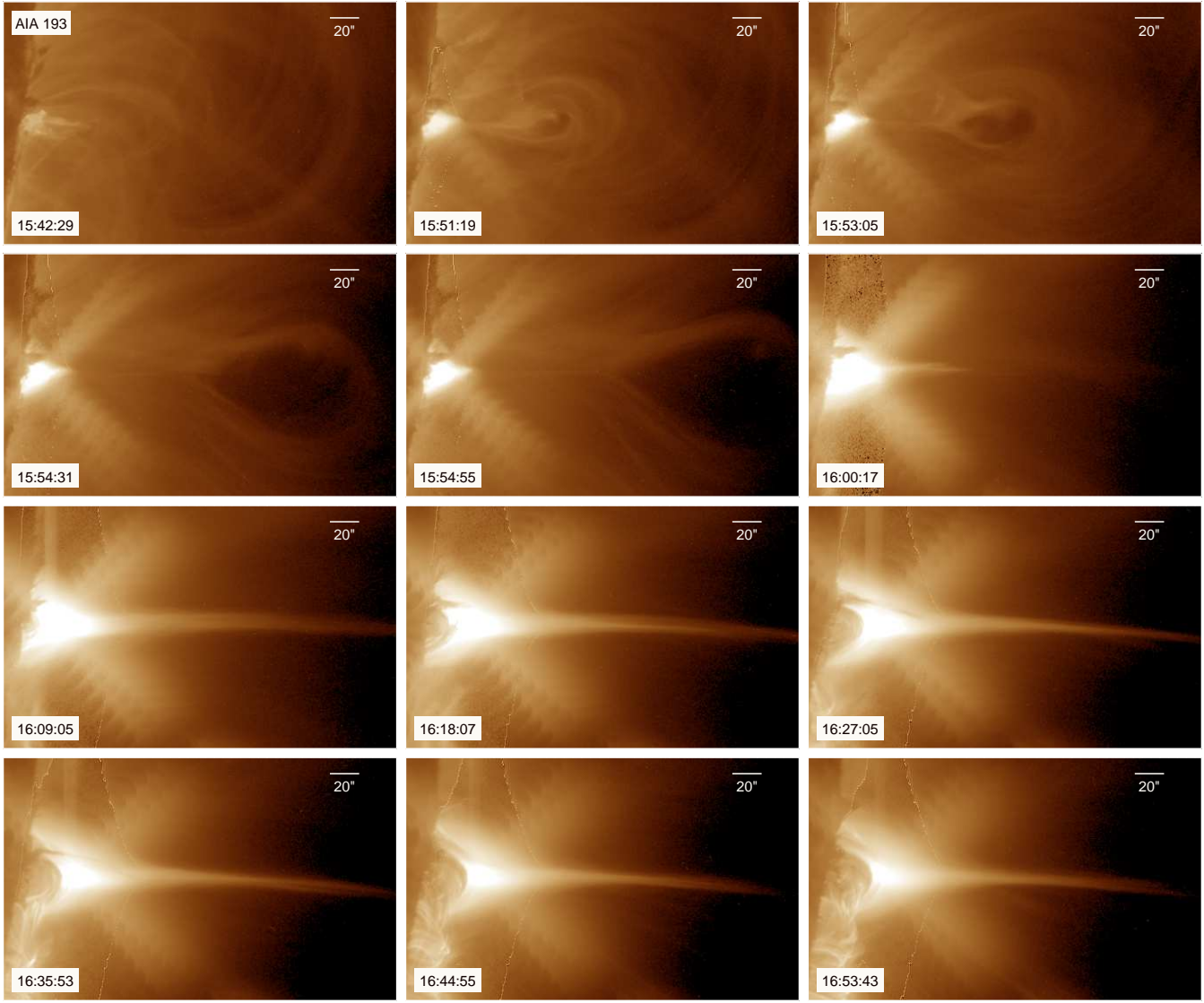}}
  \caption{A series of AIA 193\,\AA\ images for the flare. This sequence shows the eruption of the
    filament and the formation of a linear, current-sheet-like feature behind the cavity. AIA
    193\,\AA\ includes contributions from \ion{Fe}{12} and \ion{Fe}{24} and shows both the million
    degree corona and high temperature emission from the flare. The field of view shown here is
    $268\arcsec\times166\arcsec$ centered at (1068\arcsec, -143\arcsec). Note that these are
    composite images constructed from combining long and short exposure time images. The boundary
    between the two can be seen in many of the images. The wedge pattern seen in many of the images
    is the diffraction of emission from the brightest part of the flare off of the mesh supporting
    the front entrance filters.}
  \label{fig:aia_ts}
\end{figure*}

AIA is a set of multi-layer telescopes capable of imaging the Sun at high spatial resolution
(0.6\arcsec\ pixels) and high cadence (typically 12\,s). EUV images are available at 94, 131, 171,
193, 211, 304, and 335\,\AA. AIA images are also available at UV and visible wavelengths, but they
are not used in this analysis. During a flare, exposure times for the EUV channels are generally
reduced in an alternating long-short pattern. For this event, the exposure time for the 193
channel, for example, was consistently 2\,s until 15:44:31, when the exposure time of alternating
images was reduced to better accommodate the intense emission from the flare. Exposure times in 193
went as low as 4.9\,ms. The exposure times in the other EUV channels were handled similarly, except
for 171, which used a fixed exposure time throughout. For display purposes we have created
composite long-short images, which show the brightest parts of the flare from the short exposure
and the fainter emission using the long exposure. Note that all of the AIA data presented here were
downloaded from the Stanford JSOC and processed with the SolarSoftware (SSW; \citealt{freeland1998})
routine \verb+aia_prep+ using the default settings.

\begin{figure}[t!]
  \centerline{
    \includegraphics[width=0.48\textwidth]{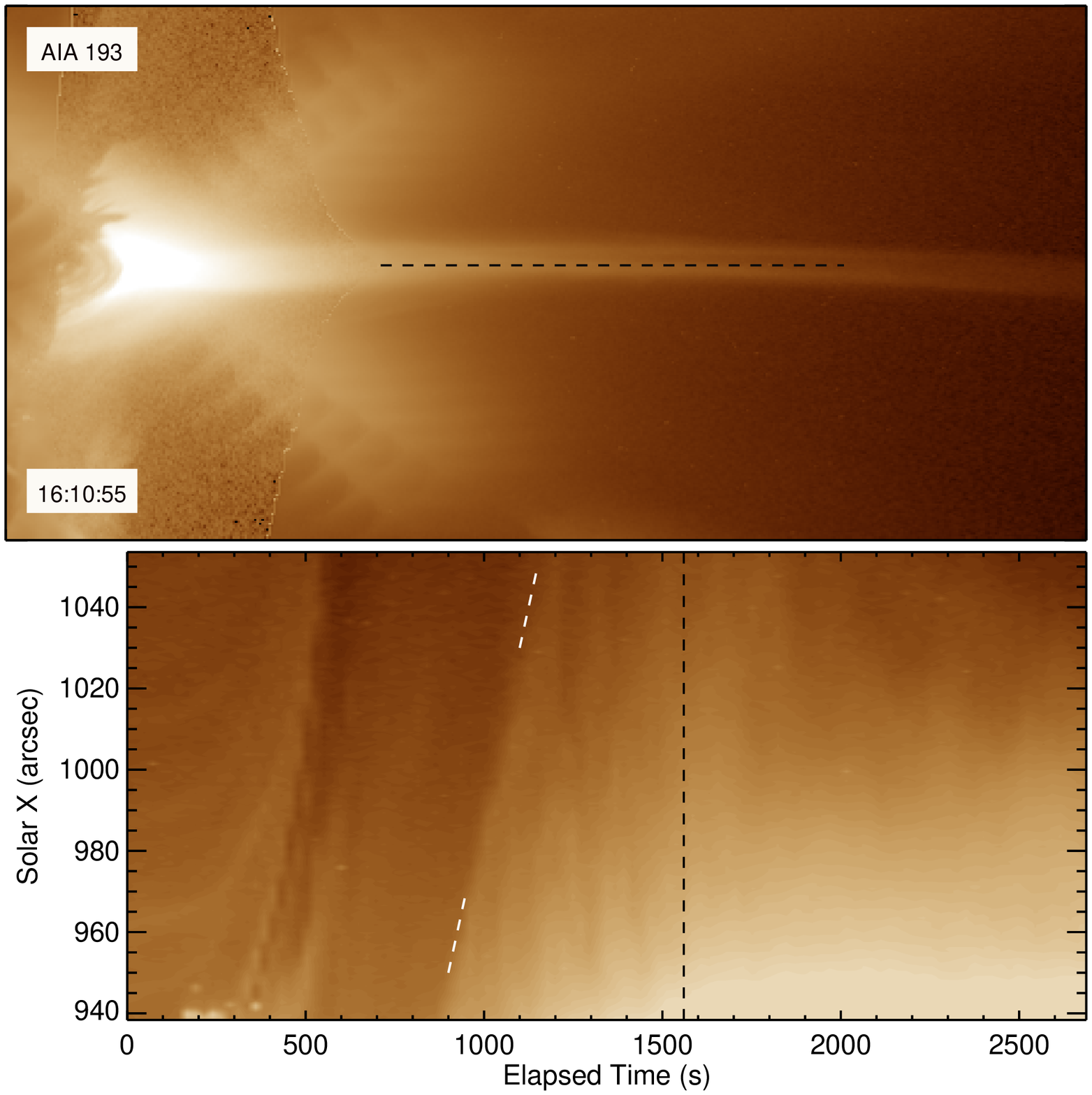}}
  \caption{Bottom: The height-time plot of intensity along the current sheet showing how it
    brightens with time. The dashed white line corresponds to a velocity of 0.4\arcsec\ s$^{-1}$,
    or 288\,km~s$^{-1}$, and represents the speed at which the intensity brightening propagates
    radially along the current sheet. Top: an example AIA 193 image showing the position of the
    ``slit.'' The time for this image is indicated by the black vertical line in the bottom
    panel. A movie version of this figure is available with the electronic version of the
    manuscript. }
  \label{fig:aia_ht}
\end{figure}

Composite 193 images showing the eruption, post-flare loop arcade, and current sheet are presented
in Figure~\ref{fig:aia_ts}. We note that the bright emission in the current sheet does not form
simultaneously.  Instead, the increase in intensity begins at the lowest heights and propagates
outward between approximately 16:00 and 16:10\,UT. To quantify this evolution we have created a
height-time plot of the intensity in a narrow region parallel to the current sheet. This is shown
in Figure~\ref{fig:aia_ht}, and yields a speed of about 288\,km~s$^{-1}$ for the propagation of the
intensity radially along the current sheet.

One instrumental artifact that is worthy of mention here is the scattering of incident photons off
of the mesh used to support the Al filters (see, for example, \citealt{poduval2013} and
\citealt{lin2001}). These filters are needed to block visible light from reaching the detector but
also must be very thin to allow for EUV emission to pass. To increase their durability they are
supported by a thin mesh. This diffraction can be seen as the wedge-pattern emanating from the
brightest emission in Figure~\ref{fig:aia_ts}.  As we will see, EIS has a similar design and also
shows this effect. The impact of this scattering on the observations will be discussed in the
Appendix.

\begin{figure*}[t!]
  \centerline{\includegraphics[angle=90, width=\textwidth]{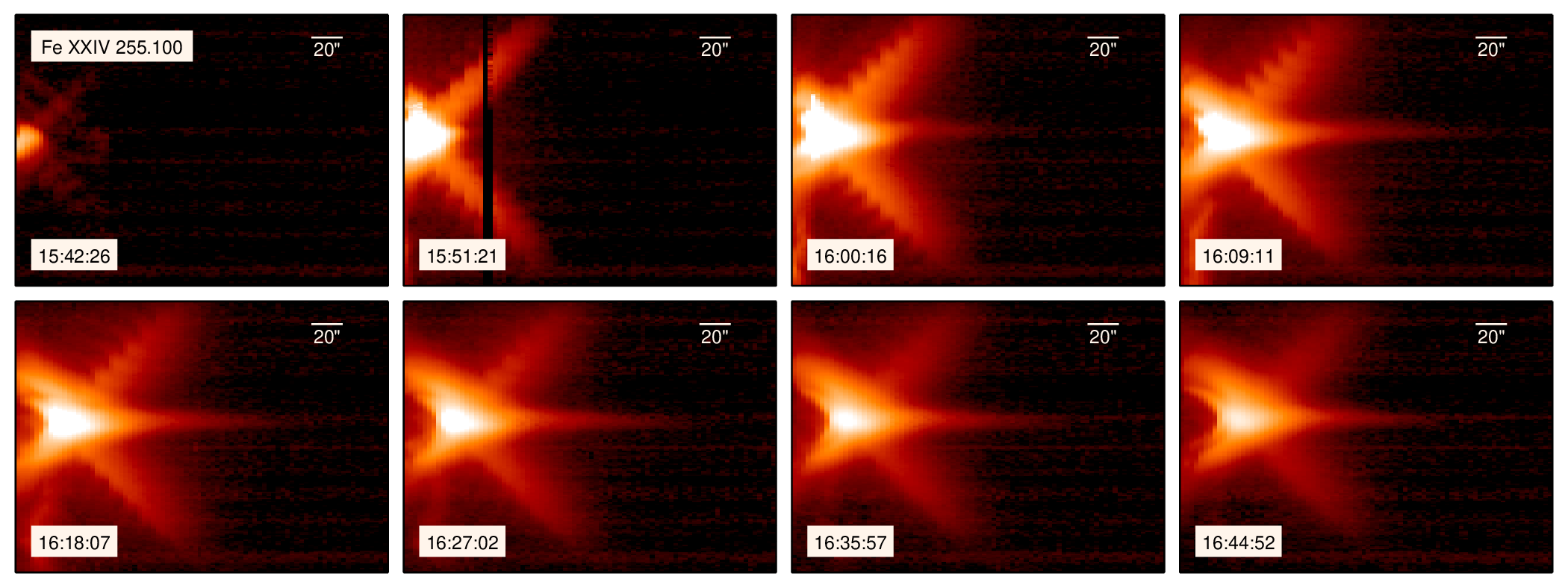}}
  \caption{A time series of EIS \ion{Fe}{24} 255.10\,\AA\ rasters. The field of view shown here is
    $240\arcsec\times174\arcsec$ centered at (1069\arcsec, -138\arcsec). Note that some of the
    brightest pixels early in the flare are saturated. As with the AIA images, the wedge pattern
    seen in many of the rasters is the diffraction of the brightest part of the flare off of the
    mesh supporting the front entrance filters.}
  \label{fig:eis_ts}
\end{figure*}

\begin{figure}[t!]
  \centerline{\includegraphics[width=0.5\textwidth]{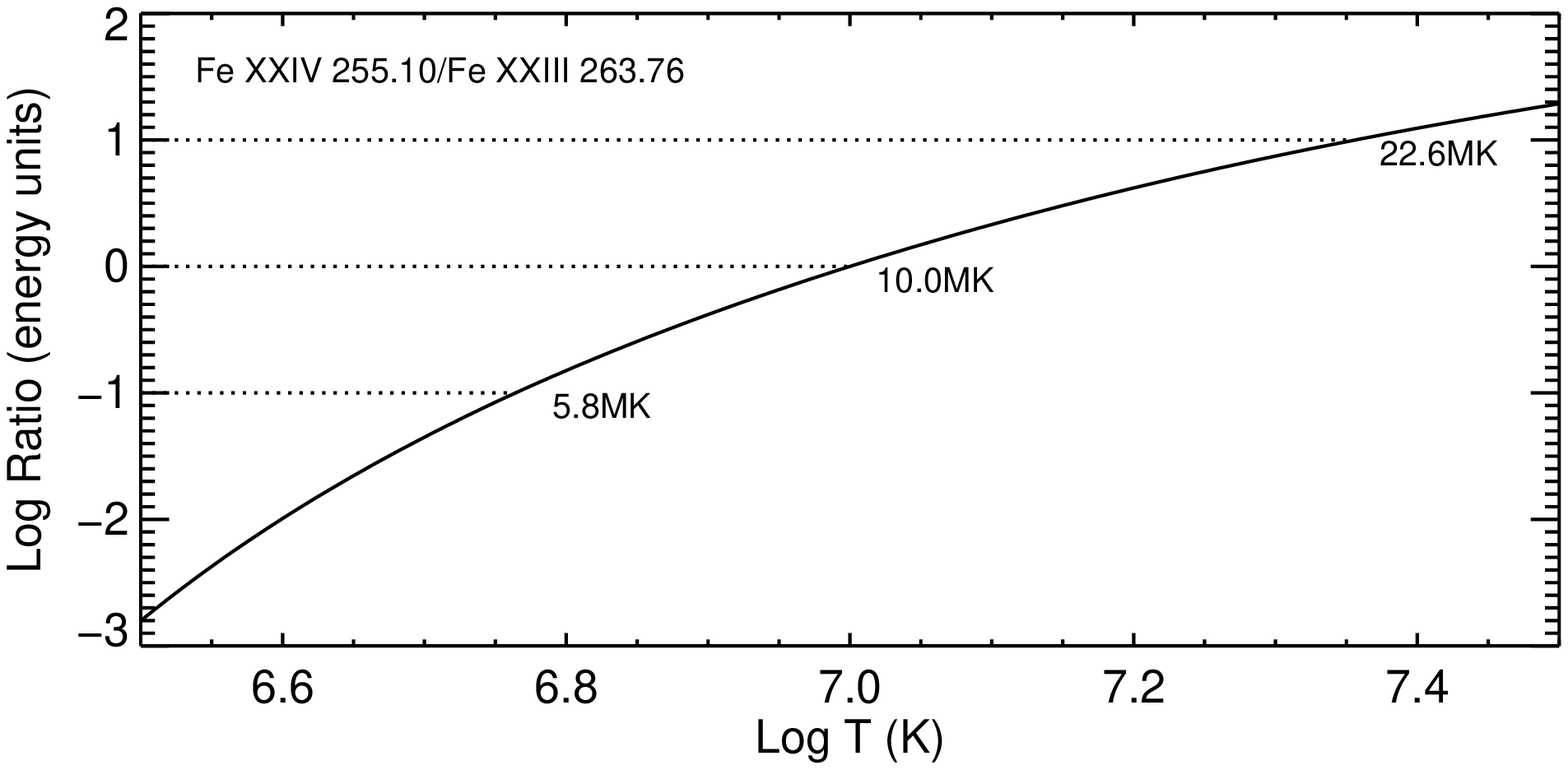}}
  \caption{The theoretical ratio of \ion{Fe}{24} 255.10\,\AA\ to \ion{Fe}{23} 263.76\,\AA\ as a
    function of temperature computed using the CHIANTI ionization fractions. The temperatures
    corresponding to ratios of 0.1, 1, and 10. are highlighted. }
  \label{fig:chianti}
\end{figure}

EIS is a high spatial and spectral resolution imaging spectrograph.  EIS observes two wavelength
ranges, 171--212\,\AA\ and 245--291\,\AA, with a spectral resolution of about 22\,m\AA\ and a
spatial pixel size of about 1\arcsec.  Solar images can be made by stepping the slit over a region
of the Sun and taking an exposure at each position. Because of telemetry constraints, EIS is often
restricted to saving only narrow spectral windows from each exposure. This restriction becomes
particularly severe when high cadence is desired. Fortunately, for this event EIS received
additional telemetry and was able to continuously run an observing sequence that saved 15 spectral
windows over a field of $240\arcsec\times304\arcsec$ using the 2\arcsec\ slit and 3\arcsec\ steps
between exposures. The exposure time at each position was fixed at 5\,s and the total time for each
raster was 535\,s. The sequence used to observe this flare began at 05:44 on September 10, 2017 and
ended at 16:53, just after the peak of the event. Observations resumed at 18:33 in a ``flare
hunting'' mode where EIS executed a very low telemetry study until a flare was detected by the
\textit{Hinode} X-Ray Telescope (XRT; \citealt{golub2007}) and then branched to the high telemetry
raster. These observations from late in the flare will be discussed in a separate paper.

All of the EIS level0 data from this time were processed using \verb+eis_prep+ with the default
settings. We fit each spectral feature of interest in each raster with a Gaussian. There are
approximately 30 emission lines available in these observations. Of primary interest here are
\ion{Fe}{24} 192.04\,\AA, 255.10\,\AA\ and \ion{Fe}{23} 263.76\,\AA. As we will show, the
temperatures in the current sheet are relatively high and the current sheet is not observed in the
lower temperature emission lines, such as \ion{Ca}{17} 192.858\,\AA.

Since the exposure time is fixed, the brightest features near the peak of the flare are
saturated. This is common for the \ion{Fe}{24} 192.04\,\AA\ line, which occurs near the peak
effective area for EIS. The effective area is smaller in the long wavelength channel and the
\ion{Fe}{24} 255.10\,\AA\ and \ion{Fe}{23} 263.76\,\AA\ lines are less affected by saturation. All
of the EIS rasters in the \ion{Fe}{24} 255.10\,\AA\ line from near the peak of the flare are
displayed in Figure~\ref{fig:eis_ts}.

\begin{figure*}[t!]
  \centerline{%
    \includegraphics[width=0.33\textwidth]{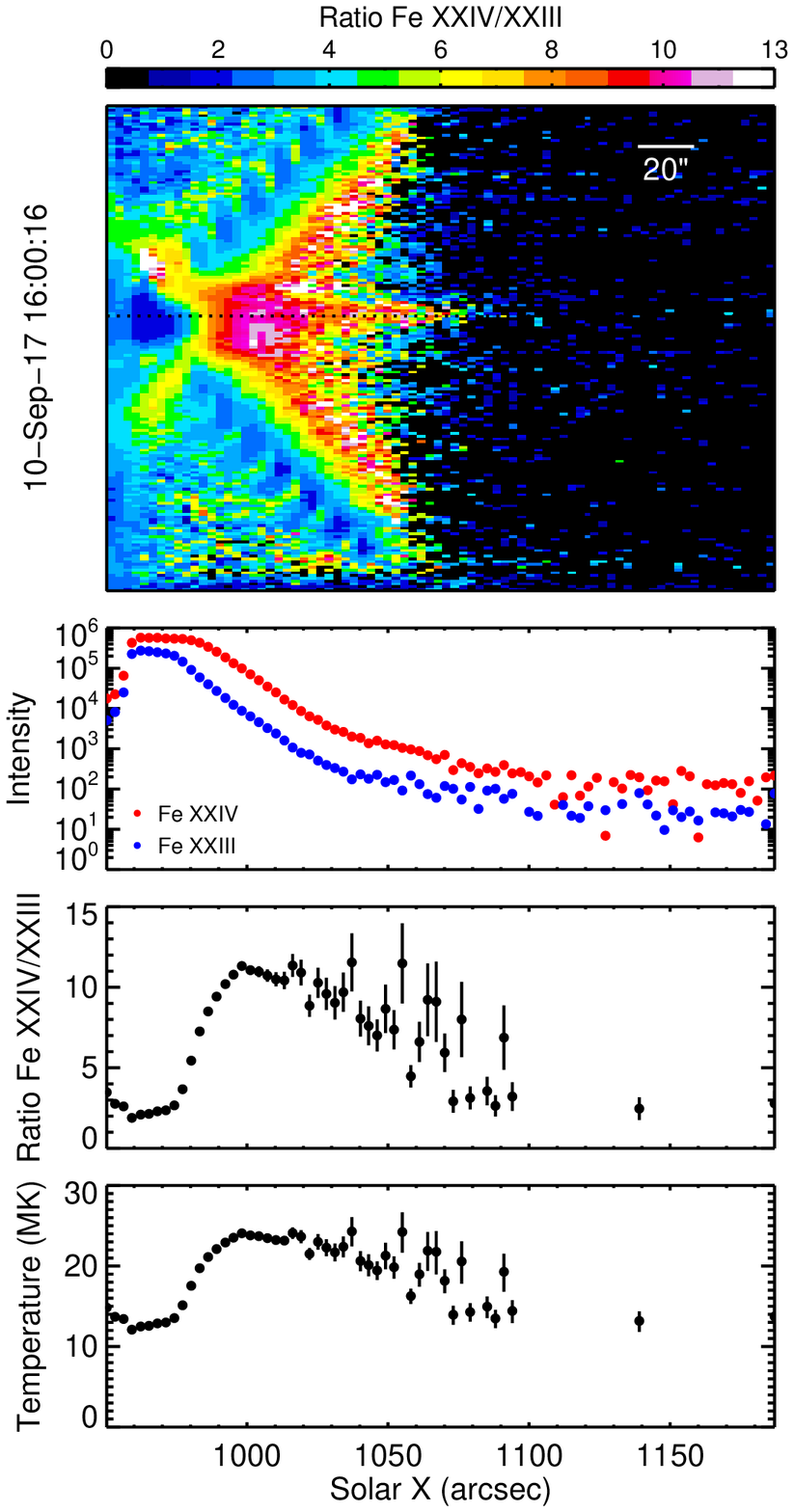}
    \includegraphics[width=0.33\textwidth]{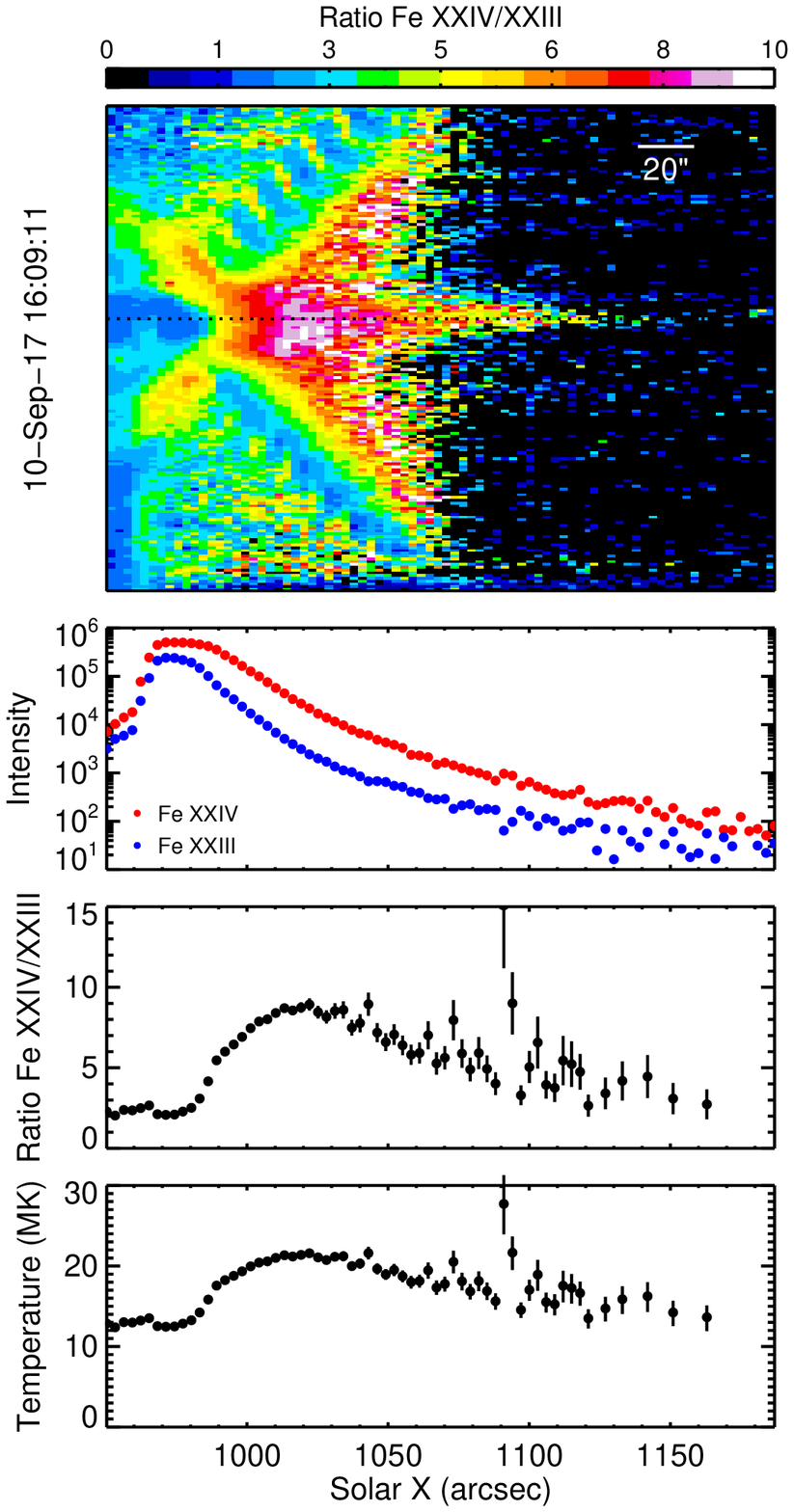}
    \includegraphics[width=0.33\textwidth]{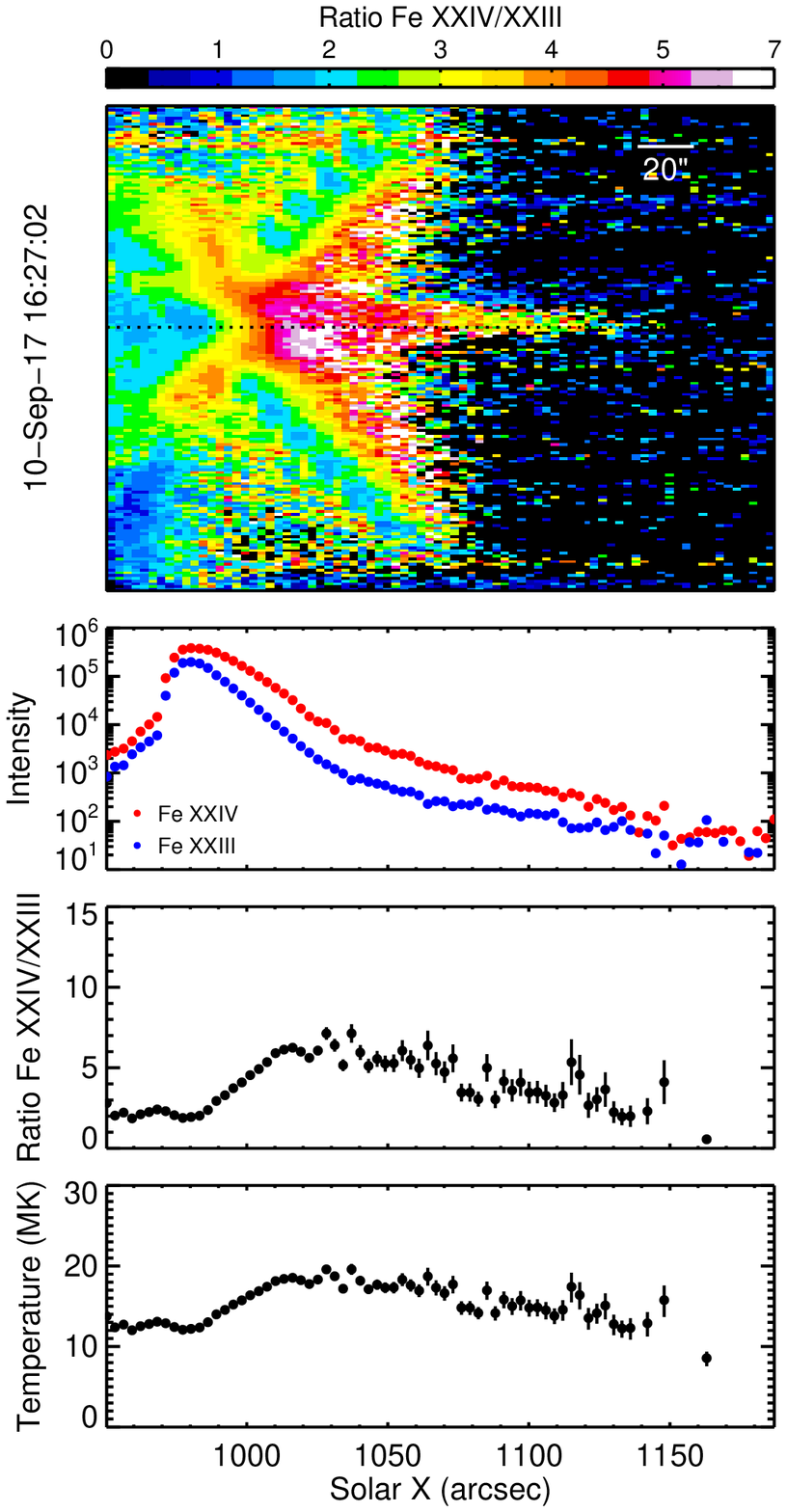}
  }
  \caption{The temperature in the flare arcade and current sheet as a function of space and time as
    determined from the EIS \ion{Fe}{24}/\ion{Fe}{23} ratio. The temperature peaks at the base of
    the current sheet and declines with height. The highest temperatures are observed early in the
    event. Note that the color scale is different for each ratio map to emphasize the relative
    differences in temperature observed at that time. The intensity, \ion{Fe}{24}/\ion{Fe}{23} ratio,
    and derived temperatures along the dotted line tracing the current sheet are shown in the lower
    panels}
  \label{fig:eis_ratio}
\end{figure*}

\subsection{Current Sheet Temperature}

We now turn to a discussion of the temperature structure of the current sheet. The ratio of the
\ion{Fe}{24} 255.10\,\AA\ to \ion{Fe}{23} 263.76\,\AA\ emission lines observed by EIS provides good
sensitivity to temperature in the 6 to 22\,MK range. This is illustrated in
Figure~\ref{fig:chianti}, where we show the theoretical ratio computed using CHIANTI v8.0.2
\citep{delzanna2015b,dere1997}.

Figure~\ref{fig:eis_ratio} shows the value of this ratio over the EIS field of view for three
different times during the flare. In each case the peak ratio, and therefore the peak temperature,
occurs at intermediate heights, essentially in the cusp region above the brightest emission in the
flare. The temperature declines with height above the limb in the current sheet. This is clearly
seen in the lower panels of Figure~\ref{fig:eis_ratio}, which show the intensity, ratio, and
temperature as a function of height in the current sheet. In each case the temperature of the
brightest emission in the flare arcade is about 12\,MK, rises to 20--25\,MK just above this region,
and then slowly declines back to about 12\,MK at the largest heights for which the ratio can be
measured. Saturation affects the brightest emission early in the flare. We can, however, infer the
temperature in this region examining the diffracted signal, which show ratios in the brightest
region of the flare of approximately 2--3. This corresponds to temperatures of 12--14\,MK.

There are significant limitations to using a single pair of emission lines to infer a
temperature. It does not, for example, yield any information on the distribution of
temperatures. Fortunately, the AIA EUV channels are also sensitive to high temperature flare
emission and we can combine the EIS and AIA observations to compute a well constrained temperature
distribution.

A detailed discussion of the AIA temperature response is given by \citet{odwyer2010}. Briefly, at
temperatures above 10\,MK the 193 channel is dominated by \ion{Fe}{24} 192.04\,\AA\ (which peaks at
$\log T = 7.25$), 131 by \ion{Fe}{21} 128.75\,\AA\ (7.05), and 94 by \ion{Fe}{18}
93.93\,\AA\ (6.85). At flare temperatures the 335 channel is affected by crosstalk from the 131
channel and has a significant contribution from \ion{Fe}{21} 128.75\,\AA\ (see
\citealt{boerner2012} Section 2.2). The 211 and 171 channels do not have a significant contribution
from emission lines formed above 10\,MK and instead are dominated by continuum emission at these
temperatures. As will be discussed in the next subsection, we can exploit this to infer the
composition in the current sheet.

\begin{figure*}[t!]
  \centerline{\includegraphics[angle=90,width=\textwidth]{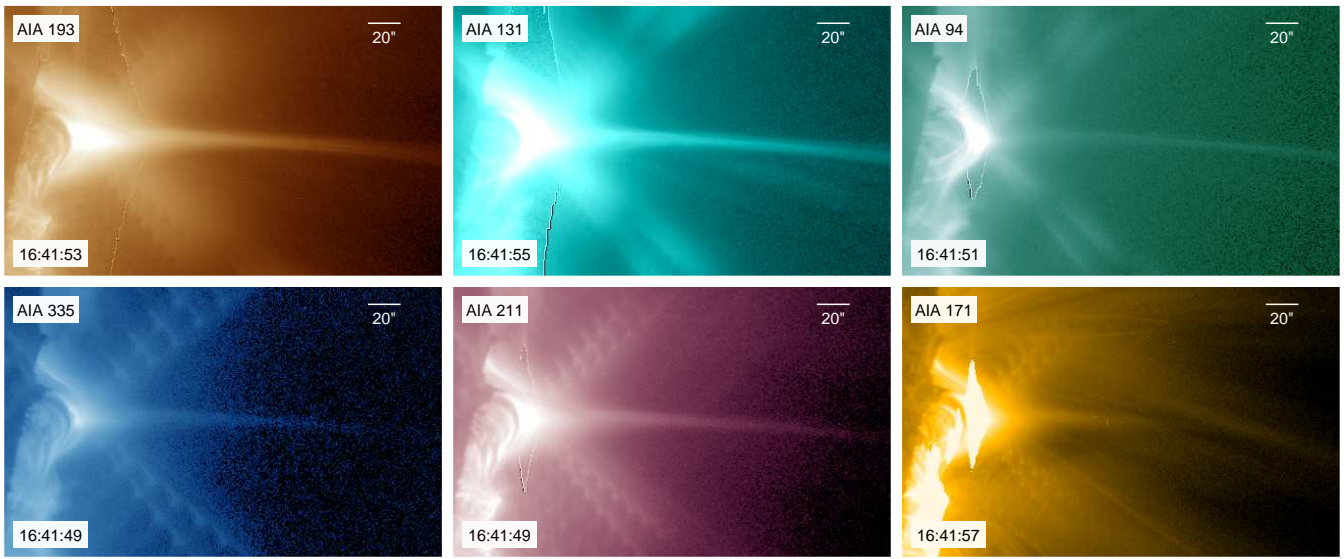}}
  \caption{AIA images near 16:41:50. The current sheet is bright in 193 (\ion{Fe}{24}) and 131
    (\ion{Fe}{21}) but relatively weak in 94 (\ion{Fe}{18}). This suggests that the temperature in
    the current sheet is relatively high (10--20\,MK) and that the distribution of temperatures is
    relatively narrow. Despite the absence of high-temperature line emission in the 211 and 171
    channels, there is some evidence of the current sheet at these wavelengths. As discussed in the
    text, this appears to be from continuum emission.}
  \label{fig:aia_set}
\end{figure*}

A set of AIA EUV images taken near 16:41:50 is shown in Figure~\ref{fig:aia_set}. The current sheet
is bright in 193 and 131 but faint in 94, suggesting that the distribution of temperatures is
relatively narrow, as there does not appear to be much emission from \ion{Fe}{18}. To quantify this
further we investigate the emission measure structure implied by the EIS and AIA observations.

\begin{figure*}[t!]
  \centerline{
    \includegraphics[angle=90,width=\textwidth]{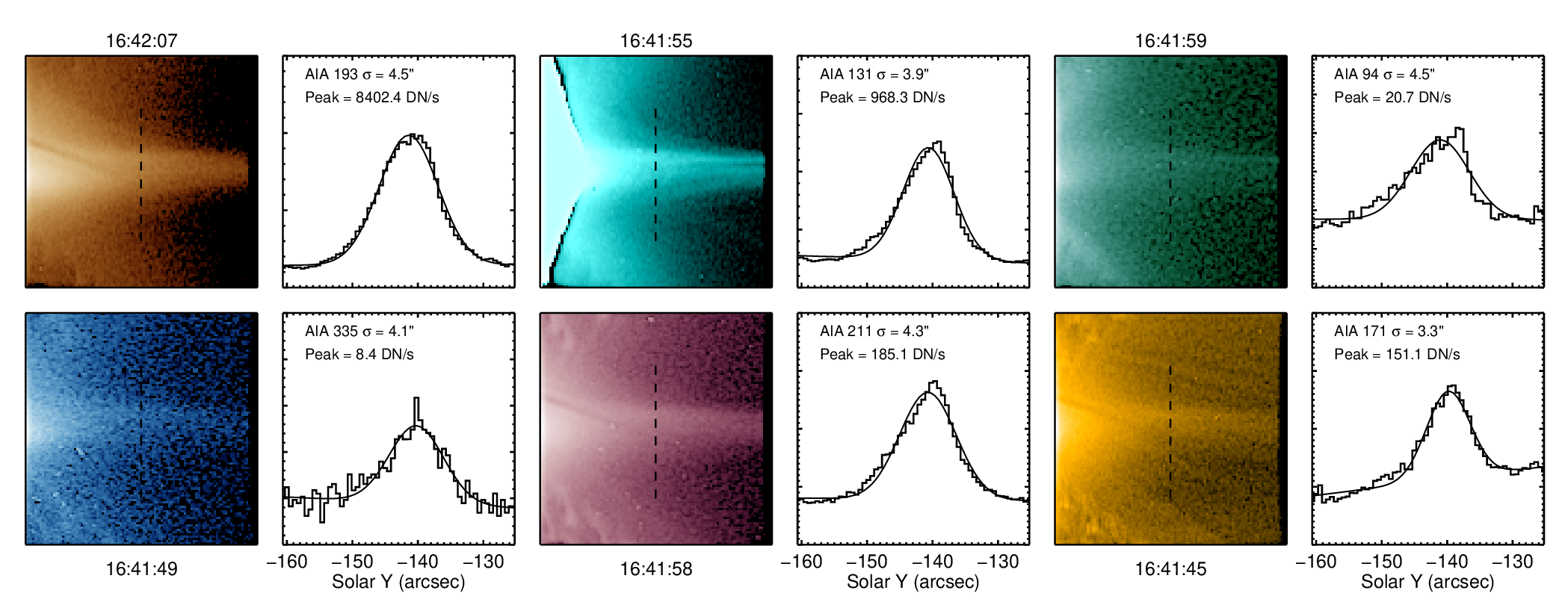}
  }
  \centerline{
    \includegraphics[width=0.345\textwidth]{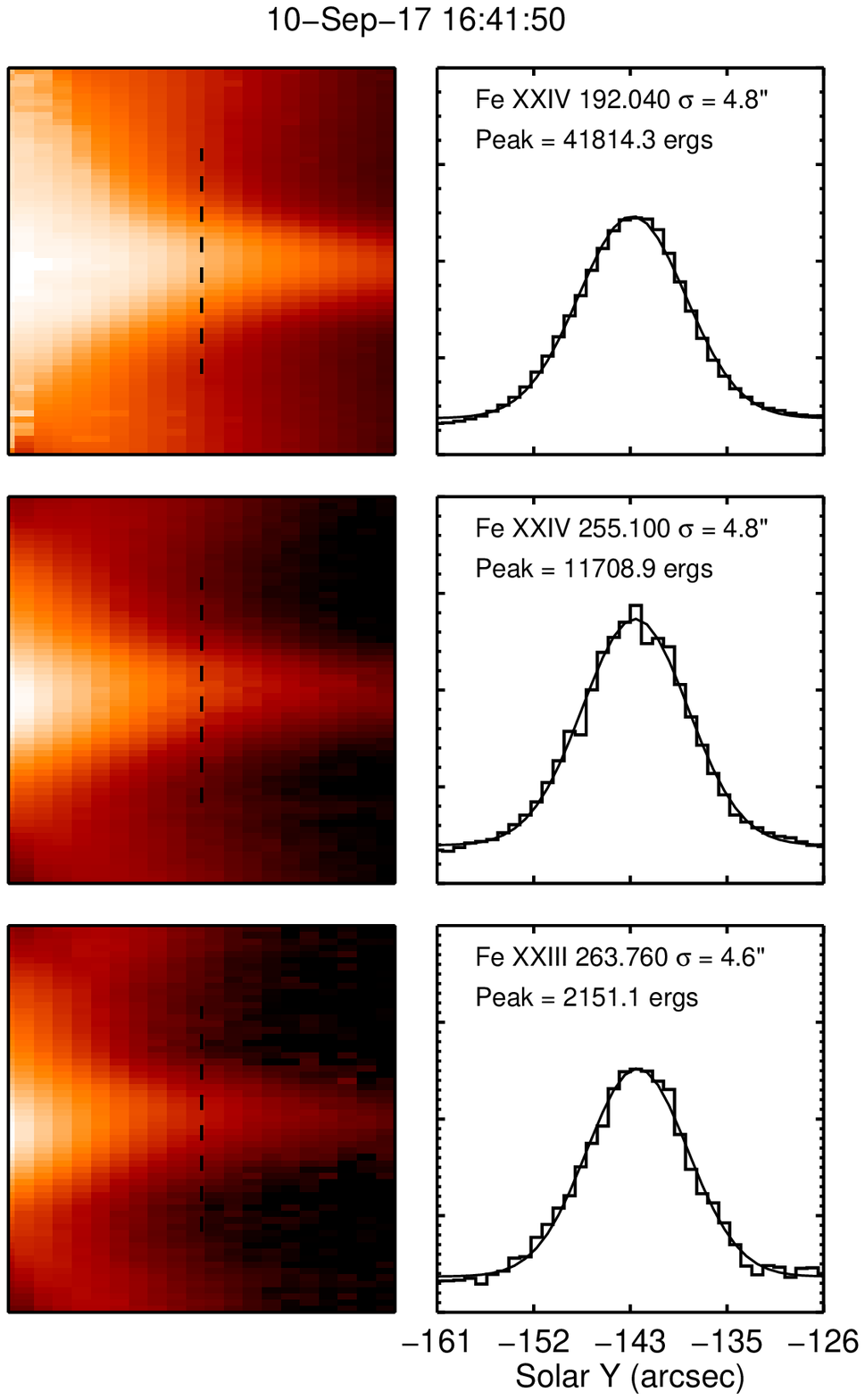}
    \includegraphics[width=0.655\textwidth]{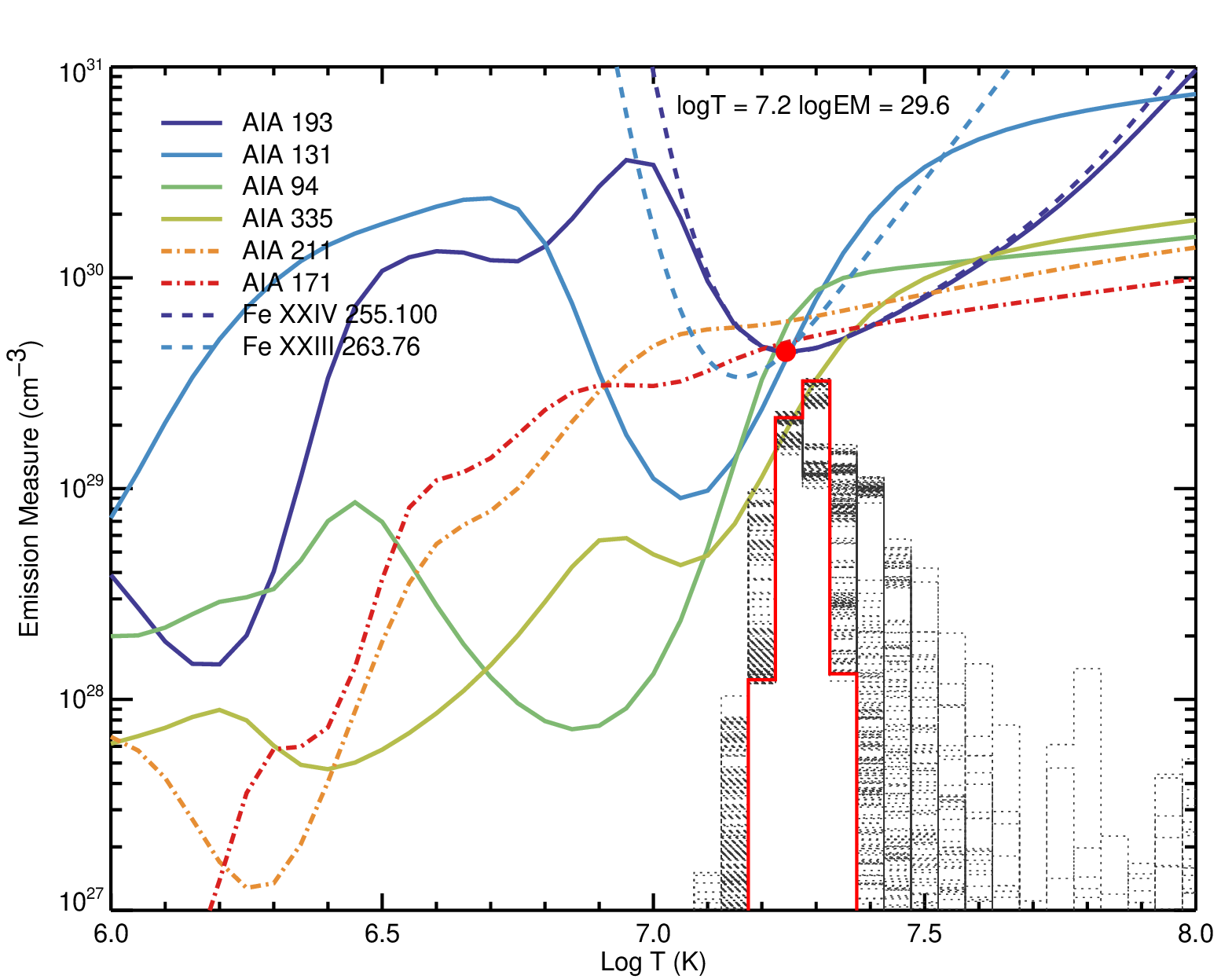}
  }
  \caption{Emission measure analysis of the current sheet observed by EIS and AIA near 16:41:50 at
    a position of (1028\arcsec,-143\arcsec). The observed intensities are approximately consistent
    with an isothermal temperature of about 16\,MK, as indicated by the red dot. An MCMC emission
    measure inversion (red line) is also consistent with a narrow temperature distribution. Also
    shown are 250 Monte Carlo MCMC calculations (thin black lines).  The paired panels each show a
    $60\arcsec\times60\arcsec$ region near the base of the current sheet and the intensities along
    a slice perpendicular to the current sheet. Intensities used in the emission measure analysis
    are derived from the peak of a Gaussian fit to data along the slice.}
  \label{fig:em_loci}
\end{figure*}

While the current sheet is clearly evident in the AIA images, there is still a contribution from
foreground and background emission from the million degree corona in these channels. To subtract
this contribution we have taken slices across the current sheet and fit the resulting intensities
with a combination of a Gaussian and second order polynomial. The peak of the Gaussian is taken as
the intensity in the current sheet at that height. This is illustrated in Figure~\ref{fig:em_loci}
where we show the intensities for a single slice across the current sheet. We have also done this
with the emission in the EIS channels. The EIS \ion{Fe}{24} and \ion{Fe}{23} emission lines
are not affected by foreground or background emission. They are, however, affected by the
diffracted signal from the mesh. This is about a 10\% effect for both lines at the lowest heights,
and the diffracted signal becomes smaller with height.

The intensities can be used to compute emission measure loci curves,
\begin{equation}
  \textrm{EM}(T) = \frac{I_{obs}}{\epsilon_\lambda(T)},
\end{equation}
where $I_{obs}$ is an observed intensity and $\epsilon_\lambda(T)$ is the corresponding emissivity
or temperature response curve. These curves form an envelope on the emission measure distribution
and are useful for investigating the temperature structure of the plasma without performing a
formal inversion. For the AIA channels we use the SSW routine \verb+aia_get_response+ with the
``evenorm'' and ``timedepend'' options to compute the temperature responses and to correct for the
decline in sensitivity over the mission. To bring the EIS \ion{Fe}{24} 255.10\,\AA\ into agreement
with AIA 193 we multiply the intensities of the EIS long wavelength lines by a factor of 1.8 (see
\citealt{delzanna2013} and \citealt{warren2014b} for a discussion of the EIS calibration). An
example set of EM loci curves is also shown in Figure~\ref{fig:em_loci}. The general confluence of
curves near a temperature of 16\,MK suggests a relatively narrow temperature distribution,
consistent with the simple interpretation of the raw images. The largest discrepancy is for the 335
channel, the origin of which is unclear. Interpreting the emission from this channel has proven
problematic in a number of different analyses (see, for example, \citealt{boerner2014}).

Finally, it is possible to use the intensities to perform a formal inversion and estimate the
differential emission measure distribution (DEM). For this we use the ``MCMC'' method as described
by \citet{kashyap1998}. This is a Bayesian method that assumes no functional form for the DEM. It
also provides estimates of the error in the DEM by recalculating the emission measure using
perturbed values for the intensities. The resulting calculation is also shown in
Figure~\ref{fig:em_loci} and confirms the narrow distribution inferred from the images.

We have computed background subtracted intensities, emission measure loci curves, and DEMs for a
number of times and positions in the flare. The resulting temperatures are consistent with those
derived from the simple EIS \ion{Fe}{24} to \ion{Fe}{23} ratio and the more detailed analysis from
Figure~\ref{fig:em_loci}. The measured temperatures are always approximately 15--20\,MK and the
temperature distributions are always narrow.

\subsection{Current Sheet Abundances}

\begin{figure*}[t!]
  \centerline{\includegraphics[angle=90,width=\textwidth]{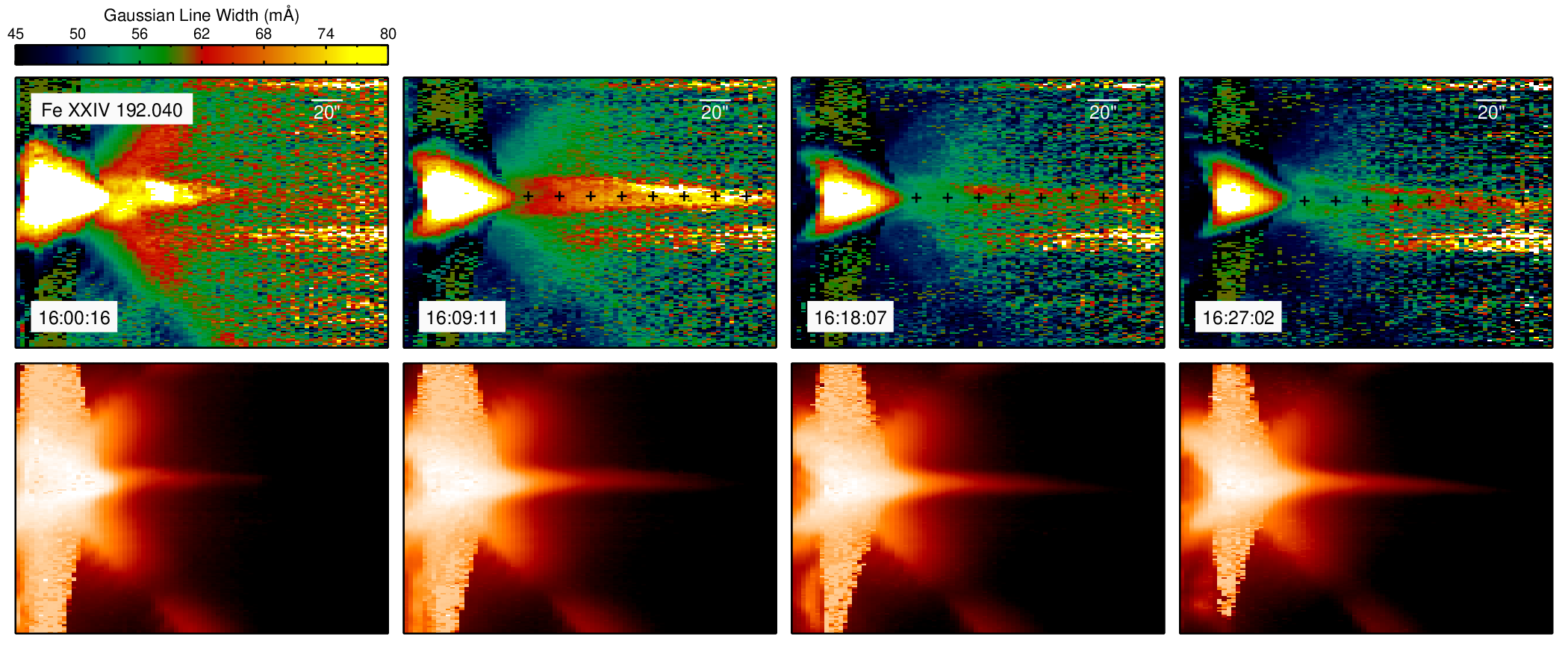}}
  \caption{Line widths and intensities derived from the EIS \ion{Fe}{24} 192.04\,\AA\ line. The
    field of view shown here is $240\arcsec\times174\arcsec$ centered at (1069\arcsec,
    -138\arcsec). The broadest profiles tend to occur early in the event and broadening appears to
    increase with height above the arcade. Line widths also decrease with time during the
    event. Non-thermal velocity calculations at selected positions (marked by the crosses) are
    shown in Figure~\ref{fig:vnt}.}
  \label{fig:widths}
\end{figure*}

Measurements of elemental abundances hold potential clues to how plasma in the solar atmosphere is
heated. It is now recognized that elemental abundances are not fixed but vary from feature to
feature and that these variations are organized by first ionization potential (FIP; for a review
see \citealt{laming2015}). There is some evidence that high temperature flare emission has
essentially a photospheric abundance \citep[e.g.,][]{warren2014}, suggesting that this plasma
originates deep in the chromosphere and is evaporated into the corona before it has time to
fractionate. Other studies indicate that the composition in flares can be complex, sometimes
showing a photospheric composition, but other times showing FIP enhancements or even an inverse FIP
effect (see \citealt{doschek2015} and \citealt{doschek2017} and references therein).

As mentioned previously, the AIA 211 and 171 channels do not include strong emission lines formed
at temperatures above 10\,MK. As can be seen in Figures~\ref{fig:aia_set} and \ref{fig:em_loci},
however, the current sheet can be seen at these wavelengths. This is likely to be from continuum
emission at these wavelengths (see \citealt{odwyer2010}), which makes the magnitude of the
intensities measured in these channels independent of the Fe abundance. We have recalculated the
AIA temperature response curves assuming a photospheric composition and confirmed this. The other
EIS wavelengths and AIA channels dominated by line emission scale linearly with the Fe
abundance. The response curves shown in Figure~\ref{fig:em_loci} are all computed assuming a
coronal composition \citep{feldman1992}. Thus the agreement of the AIA 211 and 171 emission measure
loci curves with those from the other channels indicate that the composition in the current sheet
is close to coronal. This suggests that the plasma in the current sheet originated in the corona.

\subsection{Current Sheet Turbulence}

\begin{figure*}[t!]
  \centerline{\includegraphics[angle=90,width=\textwidth]{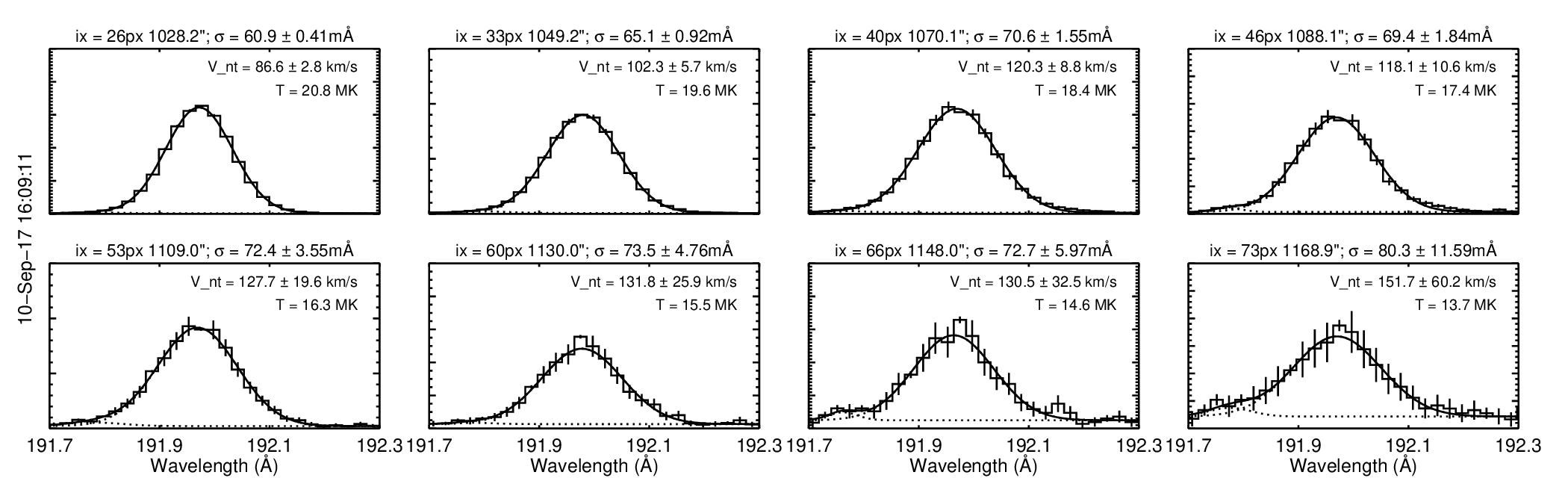}}
  \centerline{\includegraphics[angle=90,width=\textwidth]{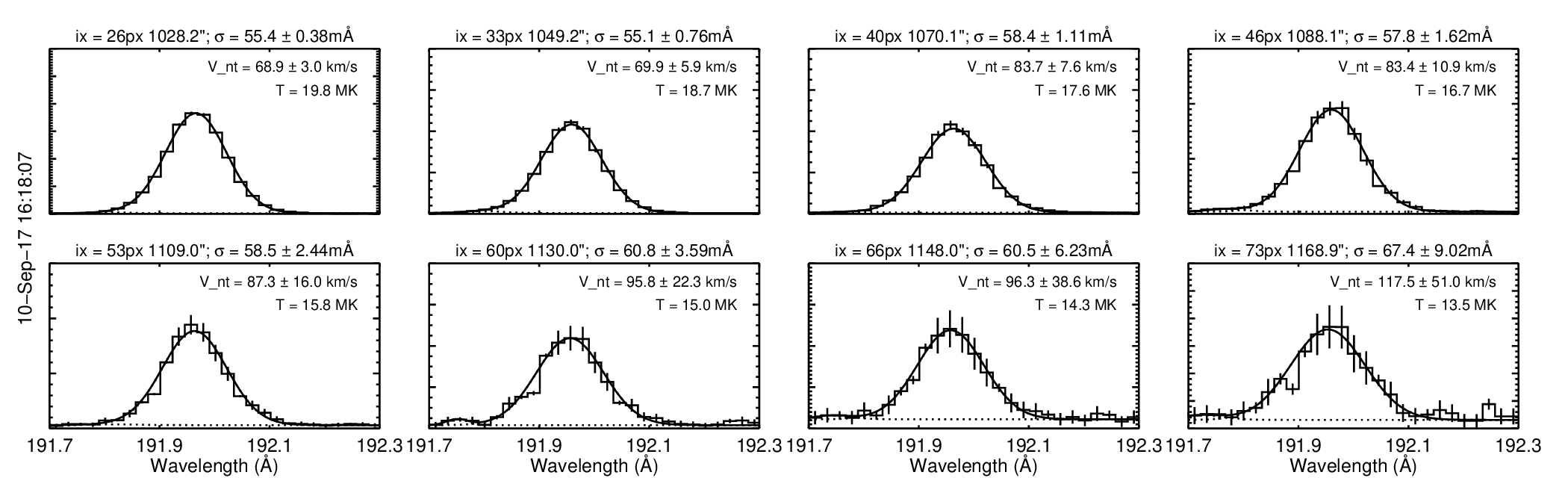}}
  \centerline{\includegraphics[angle=90,width=\textwidth]{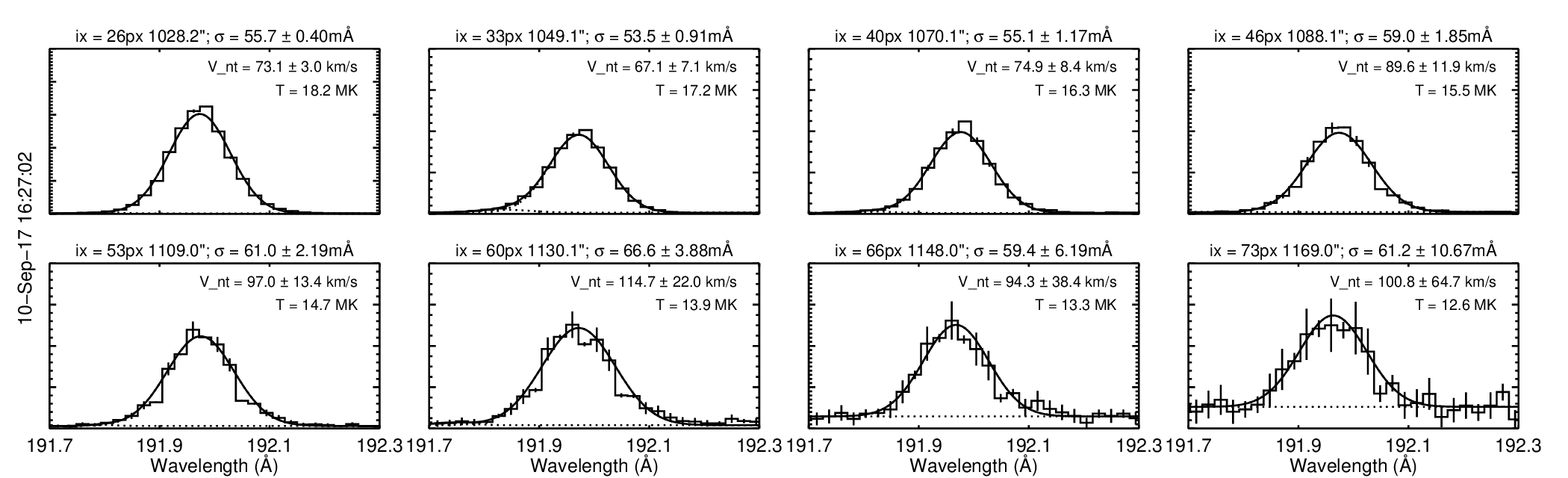}}
  \caption{EIS \ion{Fe}{24} 192.04 line profiles and non-thermal velocities as a function of height
    for various times during the early part of the event. The observed Gaussian width, observed
    temperature, and the inferred non-thermal velocity for each profile is indicated in each
    panel. The EIS instrumental width (FWHM) is approximately 70.4\,m\AA\ at this position. }
  \label{fig:vnt}
\end{figure*}

We now turn to the measurement of non-thermal velocities in the current sheet. For this we consider
\ion{Fe}{24} 192.04\,\AA. The EIS effective area is relatively high at this wavelength and the line
can generally be measured at the largest heights observed in the raster. Maps of the Gaussian line
width at several times during the flare are shown in Figure~\ref{fig:widths}. These maps clearly
show that the widest profiles are observed early in the flare, that the line widths appear to
increase with height above the limb in the current sheet, and that the line broadening diminishes
with time during the event.

The observed line width results from a combination of thermal broadening, instrumental broadening,
and non-thermal turbulent motions of the plasma. This can be expressed as
\begin{equation}
  W^2_{obs} = W_{inst}^2 + 4\ln(2)\left(\frac{\lambda}{c}\right)^2\left(v_T^2 + v_{NT}^2\right),
\end{equation}
where $\lambda$ is the wavelength of the emission, $c$ is the speed of light, and $v_T^2=2k_BT/M$
is the thermal velocity, where $k_B$ is the Boltzmann constant, $T$ is the temperature, and $M$ is
the ion mass, and $v_{NT}$ is the non-thermal velocity. To account for the temperature dependence
in the calculation we use the temperatures derived from the \ion{Fe}{24} to \ion{Fe}{23} ratio
shown in Figure~\ref{fig:eis_ratio}. Since the temperatures are somewhat noisy, we use values
interpolated from a simple exponential fit to the observed temperatures profiles.

Non-thermal velocity calculations are presented in Figure~\ref{fig:vnt} for eight positions along
the current sheet observed in the 16:09, 16:18, and 16:27 rasters. The combination of increasing
observed line width and declining temperature lead to a significant rise in the non-thermal velocity
with height in the current sheet. In the 16:09 raster, for example, we see a rise from about
87\,km~s$^{-1}$ to about 152\,km~s$^{-1}$ over the observed length of the current sheet. At the
other times the non-thermal velocities are reduced, but not as much as is suggested in the maps of
the line widths. Since the temperature is also declining with time, the non-thermal velocity does
not vanish over this period. In the 16:27 raster, for example, the values are 73\,km~s$^{-1}$ and
101\,km~s$^{-1}$ at the lowest and largest heights, respectively.
\section{Summary}

We have presented an analysis of the current sheet observed during the September 10, 2017 X8.3
flare with EIS and AIA. This analysis shows that while the temperature in the current sheet is much
higher than that of the surrounding corona, the highest temperatures are observed in the cusp of
the flare arcade and the temperature in the current sheet declines with height. We find that the
distribution of temperatures in the current sheet is relatively narrow and is locally well
described as an isothermal plasma with a temperature of 15--20\,MK. Further, the observations show
that the intensity enhancement in the current sheet begins at the lowest heights and propagates
outward at a speed of about 288\,km~s$^{-1}$.

The spatial dependence of the temperature in the flare arcade suggests that some heating occurs in
the current sheet itself, but that additional energy must be released after the field has
reconnected. One possibility is that that energy is released as field lines relax and become more
dipolar after reconnection. This process has been well observed
\citep[e.g.,][]{mckenzie1999,savage2011} and there is some indication that it leads to significant
energy release \citep[e.g.,][]{scott2016,hanneman2014,guidoni2011}. Of course, energetic particle
acceleration is also likely to play a role in energy transport during this time.

The observed intensities in the AIA 211 and 171 channels, which do not have significant
contributions from high temperature line emission, are consistent with a coronal composition. This
suggests a coronal source for the plasma in the current sheet. This is consistent with the
observations of very high temperature flare plasma observed at soft X-ray wavelengths
\citep{caspi2010}.

Finally, we observe strong non-thermal broadening of approximately 70--150\,km~s$^{-1}$ in the
current sheet. This broadening increases with height above the limb and declines with time. This is
generally consistent with the previous observations of non-thermal velocities in flares
\citep[e.g.,][]{doschek1980,antonucci1984,mariska1993}, although these spatially unresolved
measurements are likely to have been dominated by the brightest emission. Our results are also
consistent with the trend of increasing broadening with height observed in previous EIS
observations of high temperature emission just above the flare arcade \citep{doschek2014,hara2008}.

Observations of current sheets have been presented previously in the
literature. \citet{ciaravella2002}, \citet{ko2003}, \citet{ciaravella2008}, and
\citet{schettino2010} have all reported on the properties of linear structures that form in the
wake of a coronal mass ejection as observed by UVCS. Despite observing the current sheet at much
larger heights and generally much later in the event than we have, their results are largely
consistent with what we observe here. Each of the studies finds evidence for high temperature
emission, as \ion{Fe}{18}, \ion{Si}{12}, and \ion{Ca}{14} are detected. Their temperatures,
however, are generally lower than what we report here for the region just above the flare
arcade. \citet{ciaravella2002}, \citet{ko2003}, and \citet{ciaravella2008} all report FIP
enhancements of varying magnitudes in the current sheet. \citet{schettino2010} and
\citet{ciaravella2008} report strong broadening in the \ion{Fe}{18} line. \citet{seaton2017} used
AIA observations of a flare to infer temperatures in a current sheet and also find evidence for
high temperature emission. They do find much broader temperature distributions that peak at
somewhat lower temperatures than we find, but their current sheet is relatively faint and they
lacked spectroscopic data, so their inversions are not as well constrained. \citet{zhu2016} also
find temperature distributions in a current sheet that are somewhat broader and cooler than what we
observe. The current sheet emission in their event appears to be relatively intense, but lacking
spectroscopic observations the DEMs are still not likely to be well constrained.

Magnetic reconnection is widely believed to play a major role in all flaring activity and thus
extended current sheets should be present in all coronal mass ejections. It is curious then that
the literature on the physical properties of current sheets is so sparse. As mentioned in the
introduction, the obvious argument is that the current sheet should be thin and relatively tenuous
and so have a low emission measure. Our observations, however, show a current sheet that has a high
emission measure and is relatively thick. In Figure~\ref{fig:em_loci} we see an isothermal emission
measure of approximately $10^{30}$\,cm$^{-5}$, orders of magnitude higher than the background
corona \citep[e.g.,][]{warren2009}. Figure~\ref{fig:em_loci} also shows an observed width of over
3,000\,km for both EIS and AIA, much larger than the spatial resolution of either instrument (see
\citealt{brooks2012} for measurements of loops observed with EIS and AIA). The AIA studies
mentioned in the previous paragraph also show relatively wide current sheets when they are
observed. This suggests that significant heating in the current sheet may only occur under very
specific conditions.

\appendix

\section{Assymmetric PSF effects}

Previous EIS analysis has indicted that is has an asymmetric point spread function (PSF).  EIS
Software Note No. 8, available on the instrument website and distributed in SSW, reports that an
asymmetric 2D Gaussian can fit the spatial distribution of the intensity of point source candidates
in wide slit (slot) data. \citet{young2012} argue that an elliptical PSF would explain systematic
Doppler signatures observed in regions where there are strong intensity gradients, such as the
solar limb or coronal bright points in coronal holes. A similar effect has been seen in the data
from an earlier spectrometer \citep{haugan1999}.

An inclination of the spot image of a point source formed on the detector can lead to systematic
Doppler signatures in the spectra. The spill over of photons along the inclination axis results in
enhanced intensities on the blue or red side of the spectral line in pixels above and below the
center of the spot. In regions where the intensities change slowly in space, the effect from pixel
to pixel is small.  This effect, however, can be important in the presence of very strong intensity
gradients, where the spill over from brighter areas can skew the spectral line centroid
measurements in the faint areas.  The effect is simulated in the Software Note for the solar
limb. Here we discuss its effect for this flare dataset. Note that we focus on the core of the PSF,
which is relevant to the velocity signature. We have not addressed the diffracted signal from the
mesh.

\begin{figure}[t!]
   \centerline{%
      \includegraphics[bb=0 0 324 282, clip,width=0.67\textwidth]{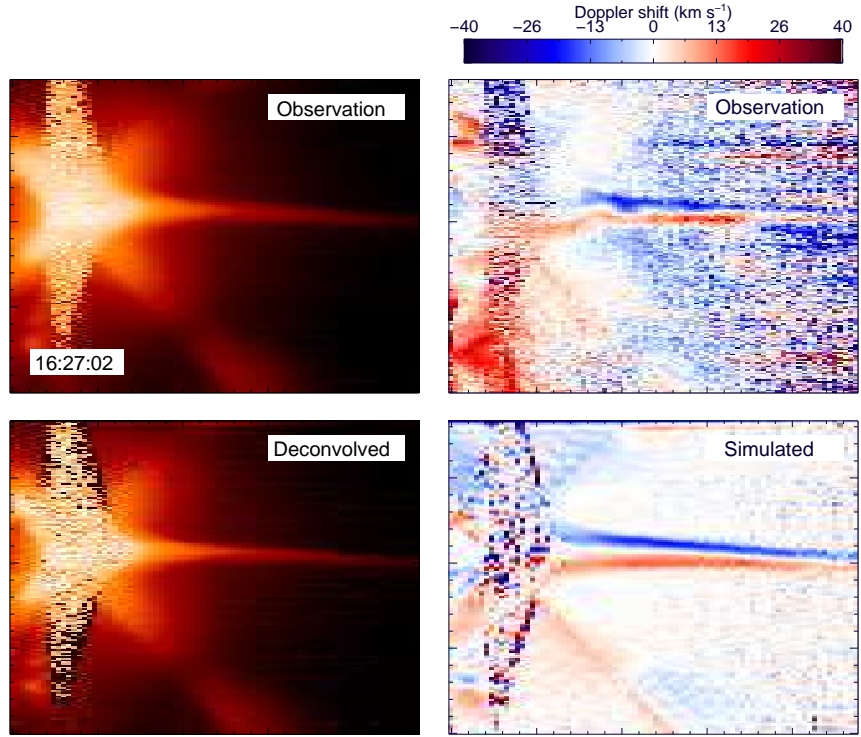}}
  \caption{EIS observes unusual velocity signatures near the current sheet, but they appear to be
  an artifact of the instrument's asymmetric PSF. Left panels: observed and deconvolved intensities
  for one of the \ion{Fe}{24} 192.04\,\AA\ rasters. These deconvolved intensities are used to
  create synthetic, zero velocity profiles that are convolved with the asymmetric PSF. Right
  panels: Observed and simulated velocity maps. The magnitude and spatial distribution of the
  simulated velocities are approximately consistent with the observations.}  \label{fig:appendix1}
\end{figure}

The current sheet in this event is a bright, elongated structure with strong intensity gradients
perpendicular to its axis, and it is a likely region for this effect to be observed.  The top row
of Figure~\ref{fig:appendix1} shows the intensity (logarithmic scale) and Doppler velocity for one
of the \ion{Fe}{24} 192.04\,\AA\ rasters. The 192.04\,\AA\ Doppler maps, which have not been
presented before in the paper, all have a distinct red and blue Doppler pattern along the current
sheet. If this were real, it would indicate systematic flows along the current sheet. As we will
see, however, it appears to be an artifact of the asymmetric PSF.

The left panel on the bottom row shows an intensity map obtained from deconvolving the observed
intensities using asymmetric Gaussian PSF [3\arcsec,4\arcsec] and the AIA library routine {\tt
aia\_deconvolve\_richardsonlucy.pro}. We assume that this is the true distribution of the flare
intensity that goes through the telescope aperture. In raster mode, every intensity column goes
through the slit and gets dispersed on the detector to form a spectrum. We used this intensity to
simulate a zero velocity spectrum centered at 192.04, assuming a thermal width at formation
temperature of the line (18 MK), and an instrumental full width half maximum of 2.5 EIS
pixels \citep{lang2006}.  We then convolved the spectra with the asymmetric PSF to see the effect
on the Doppler measurements.  We fitted each spectra with single Gaussians and determined the
centroids.  The results are shown in the velocity map in the bottom-right panel, which presents the
Doppler shift obtained from fitting the zero velocity spectra convolved with a PSF. A comparison to
the top panel reveals that systematic Doppler signatures that mimic the actual data can be
introduced by the purely instrumental effect of an inclined PSF.

It may be possible to model and subtract this instrumental effect, perhaps revealing interesting
velocity structure in the current sheet, but this is beyond the scope of the present paper. We
explored a range of widths for the PSF of 2\arcsec--5\arcsec, and the results were comparable. We
chose [3\arcsec,4\arcsec] for display purposes because it is close to the values cited in work
above, but we can not yet conclude that it provides the best fit.

Finally, we have considered the impact of asymmetries in the PSF on the temperature ratio and width
measurements. For these moments the effects appear to be small.

%% ------------------------------------------------------------------------------------------
%% --- ACKNOWLEDGMENTS ----------------------------------------------------------------------
%% ------------------------------------------------------------------------------------------

\acknowledgments This work was supported by NASA's \textit{Hinode} project. \textit{Hinode} is a
Japanese mission developed and launched by ISAS/JAXA, with NAOJ as domestic partner and NASA and
STFC (UK) as international partners. It is operated by these agencies in co-operation with ESA and
NSC (Norway). CHIANTI is a collaborative project involving George Mason University, the University
of Michigan (USA) and the University of Cambridge (UK).

%% ------------------------------------------------------------------------------------------
%% --- REFERENCES ---------------------------------------------------------------------------
%% ------------------------------------------------------------------------------------------

\end{document}